\documentclass[11pt]{article}
\usepackage{graphicx}
\usepackage{epstopdf}
\usepackage{amsmath}
\usepackage{amsfonts}
\usepackage{amssymb}

\usepackage[english]{babel}

\usepackage{cite}

\newcommand{\eq}{\begin{equation}}
\newcommand{\beq}{\begin{equation}}
\newcommand{\eeq}{\end{equation}}
\newcommand{\eqa}{\begin{eqnarray}}
\newcommand{\eeqa}{\end{eqnarray}}
\newcommand{\beqa}{\begin{eqnarray}}
\newcommand{\bea}{\begin{eqnarray}}
\newcommand{\eea}{\end{eqnarray}}

\newcommand{\mc}[1]{\mathcal{#1}}
\newcommand{\dd}{{\textrm{d}}}
\newcommand{\w}{{\wedge}}

\newcommand{\La}{{\Lambda}}

\newcommand{\p}{{\partial}}
\newcommand{\al}{{\alpha}}
\newcommand{\be}{{\beta}}

\newcommand{\si}{{\sigma}}

\newcommand{\lp}{\left(}
\newcommand{\rp}{\right)}

\newcommand{\lb}{\left[}
\newcommand{\rb}{\right]}

\newcommand{\BE}{\begin{equation}}
\newcommand{\EE}{\end{equation}}
\newcommand{\BC}{\begin{center}}
\newcommand{\EC}{\end{center}}
\newcommand{\BF}{\begin{figure}}
\newcommand{\EF}{\end{figure}}

\newcommand{\n}{\ensuremath{\nabla}}

\begin{document}

\setlength{\unitlength}{1mm}

\thispagestyle{empty}
\begin{flushright}
\small \tt
\begin{tabular}{l}
CPHT-RR023.0512\\
LPT-ORSAY 12-44
\end{tabular}
\end{flushright}
\vspace*{1.cm}

\begin{center}
{\bf \LARGE Conformally coupled scalar black holes admit a flat horizon due to axionic charge}\\

\vspace*{1.5cm}

{\bf Yannis Bardoux,}$^{1}\,$
{\bf Marco M.~Caldarelli,}$^{2,1}\,$
{\bf Christos Charmousis,}$^{1,3}\,$

\vspace*{0.5cm}

{\it $^1$\,Laboratoire de Physique Th\'eorique (LPT), Univ. Paris-Sud,\\
 CNRS UMR 8627,
 F-91405 Orsay, France}\\[.3em]

{\it $^2$\,Centre de Physique Th\'eorique (CPhT), Ecole Polytechnique,\\
 CNRS UMR 7644,
 F-91128 Palaiseau, France}\\[.3em]

{\it $^3$\,Laboratoire de Math\'ematiques et Physique Th\'eorique (LMPT), Univ. Tours,\\
UFR Sciences et Techniques,\\
Parc de Grandmont, F-37200 Tours, France}\\[.3em]

\vspace*{0.3cm} {\small\tt
yannis.bardoux@th.u-psud.fr,
marco.caldarelli@th.u-psud.fr,
christos.charmousis@th.u-psud.fr
}

\vspace*{.3cm}

\vspace{.8cm} {\bf ABSTRACT}
\end{center}

\noindent
Static, charged black holes in the presence of a negative cosmological constant and with a planar horizon are found in four dimensions. The solutions have scalar secondary hair. We claim that these constitute the planar version of the Mart\'\i nez-Troncoso-Zanelli black holes, only known up to now for a curved event horizon in four dimensions. Their planar version is rendered possible due to the presence of two, equal and homogeneously distributed, axionic charges dressing the flat horizon. The solutions are presented in the conformal and minimal frame and their basic properties and thermodynamics analysed. Entertaining recent applications to holographic superconductors, we expose two branches of solutions: the undressed axionic Reissner-Nordstr\"om-AdS black hole, and the novel black hole carrying secondary hair. We show that there is a critical temperature at which the (bald) axionic Reissner-Nordstr\"om-AdS black hole undergoes a second order phase transition to the hairy black hole spontaneously acquiring scalar hair.


\vfill \setcounter{page}{0} \setcounter{footnote}{0}
\newpage

\tableofcontents



\section{Introduction}
Topological censorship theorems, in four dimensional spacetime, severely constrain possible black hole geometries. Upon going to higher dimensions however, one can construct asymptotically flat rotating black holes of non-spherical topology, such as black rings or even more complex rotating black objects \cite{Emparan:2001wn}. In four spacetime dimensions a negative cosmological constant evades censorship and introduces an extra curvature scale, thus permitting the construction of static  black holes of zero or negative intrinsic horizon curvature.  By compactifying the transverse dimensions, arbitrary topology event horizons can be generated \cite{Lemos:1994xp,Mann:1996gj,Vanzo:1997gw}, at the cost of having  non-trivial topology at spatial infinity \cite{Galloway:1999bp}. These black holes, usually referred to as topological black holes, are easily generalized to higher dimensions \cite{Birmingham:1998nr}.

Dressing a black hole with matter is rendered difficult by the no-hair conjecture, originally enunciated by Wheeler (see \cite{Bekenstein:1998aw} for a review). The general idea behind this conjecture is that black holes, upon formation, will either eat up or expel  surrounding matter ending up as rather bald objects in their stationary state. Only charges measured at infinity remain, like electric or magnetic charge which creates a charged Reissner-Nordstr\"om black hole. Recently, axionic black holes where found for asymptotically locally anti-de Sitter spacetimes \cite{Bardoux:2012aw}. These black holes have a flat (extended) or toroidal (compact) horizon and we will refer to them as planar black holes. They are not hairy solutions since they have an associated three form charge defined at infinity. In this sense they are much like the anti-de Sitter Reissner-Nordstr\"om (RN-AdS) black holes.

Constructing hairy black holes is an interesting subject, not only for the obvious reason, i.e.~circumventing or questioning a physical conjecture, but also because they provide additional non-trivial phases of bald black holes. This, as we will see, has interesting applications to holography. A family of four dimensional hairy black holes---first discovered by Mart\'\i{}nez, Troncoso and Zanelli (MTZ)---with a cosmological constant, can be obtained with a minimally coupled scalar field \cite{Martinez:2004nb,Martinez:2006an} or with a conformally coupled scalar field \cite{Martinez:2002ru,Martinez:2005di}. The MTZ black holes do not have an independent integration constant associated to the scalar (what we would call primary hair\footnote{Recently, A. Anabal\'on et al. \cite{Anabalon:2012sn} presented a black hole with a scalar primary hair in gauged $N=8$ supergravity.}); the scalar field modifies the geometry indirectly, with its non-trivial (conformal) coupling to gravity, what is called secondary hair. In fact, this solution was first found in absence of  a cosmological constant as a counter-example attempt to the no hair conjecture \cite{Bocharova:1970,Bekenstein:1974sf}. The scalar field however diverges at the horizon \cite{Bekenstein:1975ts,Bocharova:1970} which makes its physical interpretation and stability a subject of debate \cite{Bronnikov:1978mx,McFadden:2004ni}. Its minimally coupled version is in fact singular since the conformal transformation used to pass from one frame to the other is singular well before the event horizon is reached. Furthermore, any coupling of the scalar to gravity other than conformal, renders the solutions singular \cite{Xanthopoulos:1989kb}. The cosmological constant provides an additional curvature scale, which enables to push the singularity behind the horizon of the black hole{\footnote{For discussions concerning stability however see \cite{Harper:2003wt}.}}. A positive cosmological constant does the necessary trick for a spherical horizon \cite{Martinez:2002ru} whereas a negative cosmological constant requires a hyperbolic horizon \cite{Martinez:2005di}. Unlike simple bald topological black holes, planar horizon solutions are found to be singular for the case of conformal scalars. The absence of the curvature scale for the horizon spoils altogether spacetime regularity. This is problematic, since recent applications originating from gauge/gravity duality require planar horizon black holes, e.g.~as possible non perturbative holograms/phases of layered superconductor materials. Such high temperature superconductors cannot be described using traditional perturbative BCS theory in order to calculate their dynamical transport properties, while their gravitational duals provide a promising approach to deal with the problem at strong coupling (see e.g.~the reviews \cite{Horowitz:2010gk,Hartnoll:2009sz,Herzog:2009xv}). In this paper we will circumvent the problem of finding planar solutions by adding axionic charge to the above class of conformally coupled (or minimally coupled) theories. We will show that by charging in a homogeneous fashion the planar horizon, a regular black hole solution can be constructed. Indeed, the addition of two axions homogeneously distributed on both directions of the horizon surface play the effective role of an intrinsic curvature scale rendering spacetime geometry regular \cite{Bardoux:2012aw}. The conformally coupled solutions we will present here are planar, have secondary hair and have quasi-identical properties to the hyperbolic black holes studied in \cite{Martinez:2005di,Koutsoumbas:2009pa,Martinez:2010ti}.

The hairy, charged black hole solutions constitute an additional phase to the usual bald planar RN-AdS black holes. A natural question arises which has direct application to holographic superconductors: given a set of boundary conditions which of these two black holes is thermodynamically preferred? As we lower the black hole temperature, can trivial scalars destabilize and dress the black hole? In other words, is scalar hair thermodynamically preferred below a certain
critical temperature? The picture is similar to that of hyperbolic black holes \cite{Martinez:2005di,Koutsoumbas:2009pa,Martinez:2010ti} but here there is an additional interest with respect to recent applications to holographic superconductors \cite{Horowitz:2010gk,Hartnoll:2009sz,Hartnoll:2008vx,Hartnoll:2008kx,Herzog:2009xv} since the horizon geometry is flat.

The outline of this paper is as follows. In Section~\ref{setup} we will construct the theory and the equations of motion will be derived. Then Section~\ref{sol} is devoted to the new axionic charged toroidal black holes dressed with secondary scalar hair. Its charges and thermodynamic properties are deduced using the Hamiltonian formalism in Section~\ref{sec::ham} and phase transition phenomena are analysed in Section~\ref{sec::phase}. Finally the last Section~\ref{conclusion} gives some concluding remarks and discuss possible applications to the holographic description of superconductors.

\section{Setting up the theory \label{setup}}

Consider the following action
\eqa
 \mc S &=&\frac{1}{16\pi G} \int_{\mc M} \dd^4x \sqrt{-g} \left[\left( 1-\frac{4\pi G}{3}\phi^2\right) R-2\La \right] +\nonumber \\
&-&  \int_{\mc M} \dd^4x \sqrt{-g}\left[\frac{1}{2} \left( 1-\frac{4\pi
 G}{3}\phi^2\right)^{-1}\sum_{i=1}^2  \frac{1}{3!} H_{abc}^{(i)} H^{(i)abc}  + \frac{F_{ab}F^{ab}}{16\pi} + \frac{1}{2}\p_a \phi \p^a \phi +
 \al\phi^4  \right]
\label{action}
\eeqa
where $G$ is  Newton's constant. The non-trivial coupling of the Einstein and  scalar sectors, given by $\frac{1}{12}\phi^2R$  is conformal in the presence of the standard scalar kinetic term and was used to construct the original BMBB solution \cite{Bekenstein:1975ts,Bocharova:1970}.  The overall coupling of the scalar-gravity sector, which is of course not conformal, permits us to define an effective Newton constant given by,
\eq
\label{newton}
G_\text{eff}=\frac{G}{ 1-\frac{4\pi G}{3}\phi^2} \ .
\eeq
Adding the cosmological constant with the $\phi^4$ potential, ensures again conformal coupling of the scalar, and gives the MTZ extension \cite{Martinez:2002ru}.
What is new here,  is a three form  axionic sector, represented by two exact three-forms $\mc H^{(i)} = \frac{1}{3!} H^{(i)}_{abc} \dd x^a \w \dd x^b \w \dd x^c $ ($i=1,2$),  originating from two Kalb-Ramond potentials  $\mc B^{(i)}$ such that $\mc H^{(i)} = \dd \mc B^{(i)}$. Note that they are also non-minimally coupled to the scalar field $\phi$. Hence the above action is not in the Jordan frame since matter still couples to the scalar field. The two-form  $\mc F = \frac{1}{2} F_{ab} \dd x^a \w \dd x^b = \dd \mc A$ denotes the usual Faraday tensor with its potential $\mc A$. This term is conformal in four dimensions and its coupling to the scalar is therefore trivial.

It is perhaps more illuminating to write this action  with a minimal coupling for the scalar field. Performing the following conformal transformation $\tilde{g}_{ab}=\left( 1 - \frac{4\pi G}{3}\phi^2 \right) g_{ab}$ and a  redefinition for the scalar field $\Psi = \sqrt{\frac{3}{4\pi G}}\,\text{arctanh}\left(\sqrt{\frac{4\pi G}{3}}\phi\right)$, the action \eqref{action} becomes
\BE \mc S = \int_{\mc M} \dd^4x \sqrt{-\tilde{g}}\left[\frac{\tilde{R}}{16\pi G}   - \frac{1}{2} \sum_{i=1}^2  \frac{1}{3!} H_{abc}^{(i)} H^{(i)abc}  - \frac{F_{ab}F^{ab}}{16\pi} - \frac{1}{2}\p_a \Psi \p^a \Psi - V(\Psi)  \right],
\label{minimal}
\EE
where the potential is given by
\BE V(\Psi) = \frac{\La}{8\pi G}\left[ \cosh^4\left(\sqrt{\frac{4\pi G}{3}}\Psi\right) + \frac{9\al}{2\pi\La G} \sinh^4\left(\sqrt{\frac{4\pi G}{3}}\Psi\right) \right]\,,\EE
and induces the cosmological constant $\La$. What is important to note here is that whenever the conformal factor, appearing in the effective Newton constant \eqref{newton}, is zero, the two frames are not equivalent, in other words solutions from one theory will not map in a one to one fashion to the other. We will come back to this when studying solutions to this action.

We write the equations of motion for the conformally coupled frame. Variation of \eqref{action} with respect to the metric gives,
\begin{multline}
\label{einstein}
 \left(1-\frac{4\pi G }{3}\phi^2\right)  G_{ab}  + \La g_{ab}  = 8\pi G \left(1-\frac{4\pi G }{3}\phi^2\right)^{-1} \sum_{i=1}^2 \left(\frac{1}{2}H_{acd}^{(i)} H_b^{(i)cd} - \frac{1}{12}g_{ab} H_{cde}^{(i)} H^{(i)cde} \right) \\
 + 2 G \left( F_{ac}F_b^{\ c} - \frac{1}{4}g_{ab}F_{cd} F^{cd} \right)  + 8\pi G \left(  \p_a \phi \p_b \phi - \frac{1}{2}g_{ab} \p_c \phi \p^c \phi \right) \\
 + \frac{4\pi G}{3}\left( g_{ab} \Box - \n_a \n_b \right)\phi^2 - 8\pi G \al g_{ab} \phi^4.
\end{multline}
Then varying with respect to $\phi$, $\mc A$ and $\mc B^{(i)}$, we obtain
\BE
\Box\phi = \frac{R}{6}\phi + \frac{4\pi G}{3} \phi \left( 1 - \frac{4\pi G}{3} \phi^2 \right)^{-2} \sum_{i=1}^2  \frac{1}{3!} H_{abc}^{(i)} H^{(i)abc} + 4\al \phi^3,
\label{eqphi}\EE
\BE \n_a F^{ab} = 0,  \qquad  \n_a \left[ \left( 1 - \frac{4\pi G}{3} \phi^2 \right)^{-1} H^{(i)abc} \right] =0,
\EE
respectively. An important property of the field equations stemming from the conformal coupling is the following: taking the trace of the metric equations \eqref{einstein}, and replacing it in the equation of motion for the scalar field \eqref{eqphi} gives,
\eq
\Box\phi = \frac{2}{3}\La\phi + 4\al \phi^3.
\label{boxphi}\eeq
This is also the equation of motion which emanates from the theory \eqref{action} in the absence of these axionic fields. Hence we can try to find a solution related to a hairy solution of the non-axionic theory.

\section{Axionic black holes with conformal dressing\label{sol}}

For negative cosmological constant{\footnote{For positive cosmological constant the role of the time coordinate $t$ and the radial coordinate $r$ are interchanged, thus yielding an anisotropic cosmological solution.}}, $\La = -3/\ell^2$, the theory $\eqref{action}$ admits the solution
\eq
 ds^2=-V(r)\dd t^2+\frac{\dd r^2}{V(r)}+r^2\left(\dd x^2+\dd y^2\right),\qquad
V(r)=\frac{r^2}{\ell^2}-p^2\left(1+\frac{G\mu}{r}\right)^2.
\label{hairy}\eeq
Note that the lapse function $V(r)$ is that associated to a hyperbolic black hole \cite{Martinez:2005di} even though the horizon sections defined for  $r$ and $t$ constant are flat. The crucial point here is that the axionic three-forms play the role of a horizon curvature term \cite{Bardoux:2012aw} permitting the regularity of the solution as we will see in a moment. We need two axions in order to preserve the homogeneity of  the two-dimensional horizon  \cite{Bardoux:2012aw}. If the horizon is compact then the axionic charge $p$ cannot be normalized to unity hence we keep the constant throughout the analysis. Concerning the matter sector, the solution cocktail of fields is given by\footnote{The action is invariant under the symmetry $\phi\mapsto-\phi$. We shall not recall it every time, and without loss of generality we will always choose one sign of $\phi$ to work with.},
\eq
\phi=\frac{1}{\sqrt{2\al\ell^2}}\,\frac{G \mu}{r+G \mu}, \qquad
\mc F =-\frac{q}{r^2}\,\dd t \w \dd r,  \qquad
\mc H^{(i)}=-\frac{p}{\sqrt{8 \pi G}}\left( 1 - \frac{4\pi G}{3} \phi^2 \right)\dd t \w \dd r \w \dd x^i,
\label{hairyfields}\eeq
and in particular the scalar field is real provided the coupling $\alpha$ is positive.
Moreover the constants have to satisfy the relation{\footnote{Due to the electromagnetic duality in four dimensions, one can trivially generalise the solution to include magnetic charge $g$, obtaining $q^2+g^2= p^2\mu^2G\left(\frac{2\pi G}{3\al\ell^2}-1\right)$ instead of \eqref{q} with Faraday tensor $\mc F =-\frac{q}{r^2}\,\dd t \w \dd r + g\,\dd x \w \dd y$.}}
\eq
q^2= p^2\mu^2G\left(\frac{2\pi G}{3\al\ell^2}-1\right),
\label{q}\eeq
which reduces the number of independent constants of integration to two. In addition, this constraint requires the coupling $\al$ to satisfy
\eq
0<\al\leq\frac{2\pi G}{3\ell^2},
\label{alpha}\eeq
with the upper bound being reached for uncharged $q=0$ solutions.
Furthermore, since there is no independent integration constant associated to the conformally coupled scalar field, the solution has secondary hair provided by the non-minimal coupling of the scalar with the metric and the three-forms in the action \eqref{action}. Hence, this solution describes an asymptotically locally AdS black hole with a flat event horizon with secondary hair. It is interesting to note that for $p=0$, we get the gravitational stealth solution with locally flat base manifold discovered in \cite{Martinez:2005di}.  Therefore the black hole presented here is the generalisation of this solution in the presence of mass and electric charge under the Maxwell field.

The horizon analysis and causal structure is a carbon copy of the analysis given in \cite{Martinez:2005di}. So here we repeat the important points and leave the reader to refer to the original publication for the details. The metric has a curvature singularity at $r=0$ since the scalar curvature (for example) diverges there. The scalar field on the other hand diverges at $r=-G\mu$, for $\mu$ negative. If the parameter $\mu$ is positive the black hole has a single event horizon at $r_+ = \frac{|p|\ell}{2} \left( 1 + \sqrt{1+ \frac{4 G \mu}{|p|\ell} } \right)$ whereas if $ 0 >G \mu > -\frac{|p|\ell}{4} $ there are three horizons at $   r_\pm = \frac{|p|\ell}{2} \left( 1 \pm \sqrt{1+ \frac{4 G \mu}{|p|\ell} } \right) $ and $ r_{--} = \frac{|p|\ell}{2} \left( -1 + \sqrt{1- \frac{4 G \mu}{|p|\ell} } \right) $. Here $r_{--}$ is also an event horizon and we have effectively a `black hole inside a black hole' \cite{Martinez:2005di}. The scalar field singularity lies inbetween the region bounded by $r_{--}$ and $r_-$ and there exists an extremal black hole for $r_{ext}=\frac{|p|\ell}{2} $. In other words we have,
\BE
0<r_{--}<-G \mu<r_-<r_{ext}<r_+
\label{bound1}
\EE
When $G\mu < -\frac{|p|\ell}{4}$  we still have an event horizon for $r=r_{--}$ but the scalar field explodes before reaching the horizon.  It is interesting to note that the three-form field strength remains regular all the way to the curvature singularity just like the electromagnetic field. For test matter, not coupling to the scalar field, the scalar field singularity does not influence the worldline of infalling test particles.

There is an additional particular point in $r$ where the effective Newton constant will diverge. It is given by,
$r_N=G \mu \left( -1+\sqrt{\frac{2 \pi G}{3\alpha\ell^2}} \right)$ for $\mu>0$ and $r_N=-G \mu \left( 1+\sqrt{\frac{2 \pi G}{3\alpha\ell^2}} \right)$ for $\mu<0$.
This point corresponds to a genuine curvature singularity for the solution written in the Einstein frame since the overall conformal factor of the metric is zero there. In fact $r_N<r_+$ if and only if
\eq
- \frac{\sqrt{\frac{3\al\ell^2}{2\pi G}}}{\left( 1+ \sqrt{\frac{3\al\ell^2}{2\pi G}} \right)^2 } < \frac{G \mu }{|p|\ell} < \frac{\sqrt{\frac{3\al\ell^2}{2\pi G}}}{\left( 1- \sqrt{\frac{3\al\ell^2}{2\pi G}} \right)^2 }\,.
\label{ghost}
\eeq
Therefore only for solutions verifying both inequalities do we have a black hole solution in the minimal frame \eqref{minimal}.  In this case the solution reads,
\eq
ds^2=\left(1-\frac{2\pi G}{3\alpha\ell^2}\frac{G^2\mu^2}{(r+G\mu)^2}\right)\left[-V(r)\dd t^2+\frac{\dd r^2}{V(r)}+r^2\left(\dd x^2 +\dd y^2\right)\right]
\label{bhminimal}
\eeq
with the same function given in eq.~\eqref{hairy} for $V(r)$, and the relevant matter fields defined in \eqref{minimal}.

The situation is more ambiguous and interesting in the conformal frame. There  the point $r=r_N$ corresponds to a change of sign for the entropy as we will see in the next section \cite{Martinez:2005di}. But added to that, the change of sign of the effective Newton constant will inevitably  lead to problems when looking at tensor fluctuations of this metric. One might expect the appearance of a graviton ghost in the spectrum if $r=r_N$ is outside the horizon. In any case, the divergence of the effective Newton constant reflects the failure of the theory to describe the physics of self-gravitating matter (coupling with $G_{\rm eff}$) near $r_N$.

This hairy black hole has two interesting limits. First, take the $\al\rightarrow0$, $\mu\rightarrow0$ limit, keeping the ratio $\mu^2/\al$ constant. The constraint \eqref{q} fixes this constant to be $3\ell^2q^2/(2\pi G^2p^2)$, and then the field configuration reaches the non-singular limit,
\eq
ds^2=-\lp\frac{r^2}{\ell^2}-p^2\rp\dd t^2+\lp\frac{r^2}{\ell^2}-p^2\rp^{-1}\dd r^2+r^2\lp\dd x^2+\dd y^2\rp
\label{alpha=0}\eeq
\eq
\mc F=-\frac{q}{r^2}\,\dd t\w\dd r,\qquad
\phi=\sqrt{\frac3{4\pi}}\,\frac{q}{pr}\,,\qquad
\mc H^{(i)}=-\frac{p}{\sqrt{8\pi G}}\lp1-\frac{4\pi G}{3}\phi^2\rp \dd t\w\dd r\w\dd x^i.
\eeq
This is a solution with $\al=0$, and interestingly enough the metric describes a black hole with a planar event horizon at $r_+=|p|\ell$. The scalar field is singular in the origin $r=0$, and the effective Newton constant diverges in $r_N=\sqrt G\,|q/p|$. For small Maxwell charge, $\sqrt G\,|q|<p^2\ell$, this divergence is inside the horizon, and we are left with an asymptotically locally AdS black hole solution in a theory with no quartic self-interaction term for the scalar, that is regular everywhere outside the event horizon. Taking an additional limit in which the Maxwell and axionic fields vanish ($p\rightarrow0$, $q\rightarrow 0$ keeping $q/p$ constant---or equivalently---set $p=q=0$ in the original solution), the metric \eqref{alpha=0} reduces to the AdS metric in Poincar\'e coordinates, with a non-trivial {\em stealth} scalar field that does not backreact on the geometry but is singular on the null curve $r=0$. This gravitationally stealth solution was first obtained and discussed in \cite{Martinez:2005di}.

The second limit one can take is $p\rightarrow0$ keeping $\tilde\mu=p\mu$ constant. The value of $\tilde\mu$ is set by the constraint \eqref{q}, that also requires $\al\leq2\pi G/3\ell^2$ to be satisfied. The resulting configuration of fields is,
\eq
ds^2=-\lp\frac{r^2}{\ell^2}+\frac{G_{\rm eff}q^2}{r^2}\rp\dd t^2+\lp\frac{r^2}{\ell^2}+\frac{G_{\rm eff}q^2}{r^2}\rp^{-1}\dd r^2+r^2\lp\dd x^2+\dd y^2\rp
\label{p0limit}\eeq
\eq
\mc F=-\frac{q}{r^2}\,\dd t\w\dd r,\qquad
\phi=\frac1{\sqrt{2\al\ell^2}}\,,\qquad
\mc H^{(i)}=0.
\eeq
The constant scalar field $\phi$ changes the Newton constant to a negative effective Newton constant $G_{\rm eff}$. Hence, the metric describes a black hole with a horizon for $r_+=(-G_{\rm eff}q^2\ell^2)^{1/4}>0$, and is regular up to the curvature singularity at the origin $r=0$. This is a `massless' Reissner-Nordstr\"om-AdS black hole carrying secondary hair in the form of a constant scalar field $\phi$. The reason why the electric source is dressed by an event horizon instead of being naked, is that the negative effective Newton constant makes it `anti-gravitate', flipping the sign of the term it induces in the black hole potential. However, due to the negative effective Newton constant, this solution has no minimally coupled counterpart\footnote{It is however possible to go to an Einstein frame redefining the scalar as $\Psi = \sqrt{\frac{3}{4\pi G}}\,\text{arctanh}\left(\sqrt{\frac{4\pi G}{3}}\frac1\phi\right)$. This will change the potential $V(\Psi)$, and more importantly, generate a Maxwell kinetic term of the wrong sign.} in \eqref{minimal}.

Finally, there are other solutions of the theory \eqref{action} with the same asymptotic behaviour of the fields, and we need to take them into account since they will contribute to the Euclidean path integral and be relevant for the analysis of the black hole phases.
Indeed there exist phase transitions between the hairy hyperbolic black hole of \cite{Martinez:2005di} and the hyperbolic AdS Reissner-Nordstr\"om black hole, as it is detailed in \cite{Martinez:2010ti}. Consequently it seems plausible to compare our solution with the undressed axionic Reissner-Nordstr\"om-AdS solution found in \cite{Bardoux:2012aw}. It is a solution of \eqref{action} given by,
\eq
ds^2= - V(r)\,\dd t^2 + \frac{\dd r^2}{V(r)} + r^2 \lp\dd x^2+\dd y^2\rp,\qquad
V(r)=\frac{r^2}{\ell^2}-p^2-\frac{2G\mu}{r} + \frac{G q^2}{r^2}\,,
\label{RN1}\eeq
\eq
\phi=0,\qquad
\mc F=-\frac{q}{r^2}\,\dd t\w\dd r,\qquad
\mc H^{(i)}=-\frac{p}{\sqrt{8\pi G}}\,\dd t \w\dd r\w\dd x^i.
\label{RN2}\eeq
More generally, all solutions of the minimally coupled Einstein-Maxwell theory, with gravitational coupling $G$ and two free three-form fields $H^{(i)}$ that was studied in \cite{Bardoux:2012aw},
\eq
\mc S=\frac{1}{16\pi G} \int_{\mc M} \dd^4x \sqrt{-g} \lp R-2\La- \frac{4\pi G}3 \lp H_{(1)}^2+H_{(2)}^2\rp-GF^2\rp
\label{noscalar}\eeq
are also solutions of the theory \eqref{action}.

A more general class of solutions can be found by asking the scalar $\phi$ to be constant. Equation \eqref{boxphi} requires its constant value to be set to
\eq
\phi=\frac{1}{\sqrt{2\al\ell^2}}.
\label{constscalar}\eeq
Then, the field equations for the remaining fields reduce to the field equations of the theory
\eq
 \mc S =\frac{1}{16\pi G_{\rm eff}} \int_{\mc M} \dd^4x \sqrt{-g} \left( R-2\La
-\frac{4\pi G_{\rm eff}}3 \left( 1-\frac{4\pi G}{3}\phi^2\right)^{-1}\lp H_{(1)}^2+H_{(2)}^2\rp-G_{\rm eff}F^2\rp,
\eeq
and hence any solution of \eqref{noscalar} leads to a new solution with constant scalar field \eqref{constscalar} by replacing $G\mapsto G_{\rm eff}$ and rescaling the three-form fields accordingly. Through this map, the axionic RN-AdS solution \eqref{RN1}--\eqref{RN2} leads to,
\eq
ds^2= - V(r)\,\dd t^2 + \frac{\dd r^2}{V(r)} + r^2 \lp\dd x^2+\dd y^2\rp,\qquad
V(r)=\frac{r^2}{\ell^2}-p^2-\frac{2G_{\rm eff}\mu}{r} + \frac{G_{\rm eff} q^2}{r^2}\,, \label{phiRN1}
\eeq
\eq
\phi=\frac{1}{\sqrt{2\al\ell^2}},\qquad
\mc F=-\frac{q}{r^2}\,\dd t\w\dd r,\qquad
\mc H^{(i)}=-\frac{p}{\sqrt{8\pi G}}\,\dd t \w\dd r\w\dd x^i.
\label{phiRN2}
\eeq
When $p=\mu=0$, this solution reduces to the solution \eqref{p0limit}, connecting this family of solutions to the family of hairy black holes \eqref{hairy}. Nevertheless, like \eqref{p0limit}, this solution has a negative effective Newton constant.

\section{Hamiltonian analysis: charges and thermodynamic properties\label{sec::ham}}

In the Euclidean approach, the partition function for a thermodynamical ensemble is identified with the Euclidean path integral in the saddle point approximation around the Euclidean section of the classical solution \cite{Gibbons:1976ue}. The period $\be$ of the imaginary time $\tau=-it$ is identified with the inverse temperature, and the free energy (or `thermodynamic potential') $W$ is related to the Euclidean action by $I_E=\be W$.

For simplicity, and without loss of generality, we will restrict the configuration space to a static metric, with planar symmetry on the constant $(t,r)$ sections. We consider therefore the family of configurations with Euclidean metric of the form
\eq
ds_E^2 = N^2(r) f(r)\, \dd \tau^2 + \frac{\dd r^2}{f(r)} + r^2 \left( \dd x^2 + \dd y^2 \right),
\label{emetric}\eeq
and fields satisfying the ansatz
\eq
\phi = \phi(r), \qquad
\mc A = A(r)\,\dd t, \qquad
\mc B^{(i)} = B_i(r)\,\dd t \w \dd x^i.
\eeq
While the planar symmetry is broken by the $\mc H^{(i)}$ fields taken on their own, the solutions of the equations of motion in our ansatz will be such that they back react isotropically on the horizon metric \cite{Bardoux:2012aw}.

Asymptotically, we impose that the fields behave as,
\eq
\begin{array}{l@{\qquad}l}
\displaystyle f(r)=\frac{r^2}{\ell^2}+f_0+\frac{f_1}r+\mc O\lp\frac1{r^2}\rp,
&\displaystyle N(r)=N_0+\mc O\lp\frac1{r^2}\rp,
\vphantom{\lp\frac1{r^2_{p_p}}\rp}\\
\displaystyle A(r)=A_0+\frac{4\pi N_0\pi_A}r+\mc O\lp\frac1{r^2}\rp,
&\displaystyle B_i(r)=-2rN_0\pi_{B_i}+B^0_i+\mc O\lp\frac1r\rp,
\vphantom{\lp\frac1{r^2_{p_p}}\rp}\\
\displaystyle \phi(r)=\frac{\phi_1}{r}+\frac{\phi_2}{r^2}+\mc O\lp\frac1{r^3}\rp,
\end{array}
\label{falloff}\eeq
Here, $\{f_0, f_1,N_0,\phi_1,\phi_2,A_0,\pi_A,B^0_i,\pi_{B_i}\}$ is a set of constants that determines the asymptotic falloff of the fields, fulfilling the additional constraints
\eq
f_0=-16\pi G\,\lp\pi_{B_1}^2+\pi_{B_2}^2\rp,\qquad
\phi_2=-\sqrt{2\al\ell^2}\,\phi_1^2,
\label{constraints}\eeq
that one obtains by solving asymptotically the equations of motion. The expansions of $f$ and $N$ ensure that the spacetime is asymptotically locally AdS, and we can set $N_0=1$ without loss of generality, by rescaling the time coordinate.

To ensure the presence of a black hole in the spacetime \eqref{emetric}, we require $f(r)$ to vanish in $r=r_h$, $r_h$ being its largest root. In the Euclidean formalism, the horizon acts as a second boundary, close to which we have to restrict the field behavior. We have thus
\eq
f(r)=f'(r_h)(r-r_h)+\mc O\lp(r-r_h)^2\rp.
\eeq
Requiring the absence of a conical singularity on the bolt we connect the values of the fields on the horizon to the inverse temperature according to
\eq
N(r_h)f'(r_h)=\frac{4\pi}\be.
\label{conical}\eeq
Finally, we require the other matter fields $\phi(r)$, $A(r)$ and $B^{(i)}(r)$ to assume finite limits on the horizon, that we will call $\phi_h$, $A_h$ and $B^{(i)}_h$ respectively.

These boundary conditions define the configuration space for our Hamiltonian analysis, and we emphasize that both solutions of interest, the hairy black hole \eqref{hairy}--\eqref{hairyfields} and the axionic Reissner-Nordstr\"om-AdS \eqref{RN1}--\eqref{RN2}, belong to this space.

On this minisuperspace and with the boundary conditions we just defined, the Euclidean action reads
\eq
\mc S_E = \si\be \int_{r_h}^{\infty}\!dr\lp N\mc H-A\pi_A'-2B_1\pi_{B_1}'-2B_2\pi_{B_2}'\rp + \mc Q
\eeq
where $\mc Q$ is the boundary term and $\sigma$ is the area of the transverse sections, coming from the integration along the $x$ and $y$ directions. While $\sigma$ is formally infinite, we can always compactify the horizon on a two-torus and make it finite. The Hamiltonian constraint reads,
\eqa
\mc H=2\lp1-\frac{4\pi G}3\phi^2\rp\lp\pi_{B_1}^2+\pi_{B_2}^2\rp+\frac{2\pi}{r^2}\pi_A^2
+\frac{r^2}{8\pi G}\lb\lp1-\frac{4\pi G}3\phi^2\rp\lp\frac{f'}r+\frac f{r^2}\rp+\La\rb\nonumber\\
+\frac{r^2}6\lb
f\phi'^2-\lp f'+\frac{4f}r\rp\phi\phi'-2f\phi\phi''+6\al\phi^4
\rb.\qquad\qquad\qquad\qquad\qquad\quad
\eeqa
Here $\pi_A$ and $\pi_{B_i}$ are the $r$ components of the conjugate momentum to $A_\mu$ and the $(ri)$-components of the conjugate momentum to $B^{(i)}_{\mu\nu}$ respectively{\footnote{As we will see, the conjugate momenta are constant and assume values identical to the constants $\pi_A$ and $\pi_{B_i}$ defined above \eqref{constraints}, and we hence keep the same notation.}}. Within our ansatz they read,
\eq
\pi_A=-\frac{r^2A'}{4\pi N},\qquad
\pi_{B_i}=-\frac{B_i'}{2N\lp1-\frac{4\pi G}3\phi^2\rp}.
\eeq
To have a well-defined variational problem, the action must be a differentiable functional of the canonical variables $\{A,\pi_A,B_i,\pi_{B_1},f,\phi\}$ on the configuration space defined by the boundary conditions \eqref{falloff}--\eqref{conical} (see \cite{Regge:1974zd}). Hence, when performing the total variation $\delta S_E$ of the action, the $\delta\mc Q$ variation of the boundary term should cancel precisely the boundary terms induced by the variation of the bulk term, yielding the equations of motion and the constraints of the theory. It follows that the variation of the boundary terms must be given by,
\eqa
\delta\mc Q=\si\be\lb
A\,\delta\pi_A+2B_1\,\delta\pi_{B_1}+2B_2\,\delta\pi_{B_2}
-\frac{rN}{8\pi G}\lp1-\frac{4\pi G}3\phi^2-\frac{4\pi G}3r\phi\phi'\rp\delta f
\right.\qquad\nonumber\\
\left.
-\frac{r^2N}6\lp4f\phi'+f'\phi+\frac{2N'}Nf\phi\rp\delta\phi
+\frac{r^2N}3f\phi\,\delta\phi'
\rb_{r_h}^\infty.
\eeqa
On the horizon, the variation of the fields is given by
\eq
\delta f|_{r_h}=-f'(r_h)\delta r_h,\qquad
\delta \phi|_{r_h}=\delta\phi(r_h)-\phi'(r_h)\delta r_h\,.
\eeq
We regularise the boundary term by taking the asymptotic boundary at $r=R$. Then, defining the effective Newton constant on the horizon $G_h$,
\eq
G_h=\frac{G}{1-\frac{4\pi G}3\phi^2(r_h)},
\eeq
and considering the falloff conditions \eqref{falloff} on the fields, we obtain
\eqa
\delta\mc Q=\beta\si\,\Phi\,\delta\pi_A+2\be\si\,\Psi_1\,\delta\pi_{B_1}+2\be\si\,\Psi_2\,\delta\pi_{B_2}
-\delta\lp\frac{\si r_h^2}{4G_h}\rp
-\frac{\be\si}{8\pi G}\,\delta f_1\qquad\qquad\qquad\nonumber\\
-\be\si\,R\lp4\pi_{B_1}\delta\pi_{B_1}+4\pi_{B_2}\delta\pi_{B_2}+\frac{\delta f_0}{8\pi G}\rp
+\frac{\be\si}{3\ell^2}\lp2\phi_2\delta\phi_1-\phi_1\delta\phi_2\rp
+\mc O\lp\frac1R\rp.
\eeqa
Here we defined the potential differences $\Phi$ and $\Psi_i$ between the spatial infinity and the event horizon for the Maxwell and axionic fields as
\eq
\Phi=A_0-A(r_h),\qquad
\Psi_i=B^0_i-B_i(r_h).
\eeq
The variations appearing here are not all independent, and by virtue of \eqref{constraints} they are related by
\eq
\delta f_0=-16\pi G\,\delta\lp\pi_{B_1}^2+\pi_{B_2}^2\rp,\qquad
\phi_1\delta\phi_2=2\phi_2\delta\phi_1.
\eeq
Therefore, the linearly diverging term in $\delta\mc Q$ vanishes, and we can safely remove the cutoff by sending $R$ to infinity,
\eq
\delta\mc Q=\beta\si\,\Phi\,\delta\pi_A+2\be\si\,\Psi_1\,\delta\pi_{B_1}+2\be\si\,\Psi_2\,\delta\pi_{B_2}-\delta\lp\frac{\si r_h^2}{4G_h}\rp
-\frac{\be\si}{8\pi G}\delta f_1.
\label{deltaQ}\eeq
With the boundary conditions we are using, we are keeping $\beta$, $\Phi$ and $\Psi_{i}$ fixed, and therefore $\delta\mc Q$ is a total variation, from which we can readily deduce the boundary term $\mc Q$. This proves that given $\mc Q$, the Euclidean action is a differentiable functional on the configuration space defined by our boundary conditions, and the Euclidean solutions of the theory \eqref{action} are extrema of $S_E$. The vanishing of the variations with respect to the `Killing lapse' $N$ imposes the Hamiltonian constraint $\delta\mc H=0$. The potentials $A$ and $B_i$ are also Lagrange multipliers, that establish that their conjugate momenta $\pi_A$ and $\pi_{B_i}$ are constants of motion, in agreement with Gauss' law. Finally, the $f$ and $\phi$ variations yield the remaining Einstein equation, and the equation for the scalar field.

Since the bulk terms of the action are pure constraints, they vanish on-shell, and the Euclidean action evaluated on solutions of the theory is completely determined by the value of $\mc Q$,
\eq
\mc S_E = -\frac{\mc A_h}{4G_h}-\frac{\be\si}{8\pi G}f_1+\be\si\lp
\Phi\pi_A+2\Psi_1\pi_{B_1}+2\Psi_2\pi_{B_2}\rp,
\eeq
where we have introduced the horizon area $\mc A_h =\sigma r_h^2$ and $G_h$ is the effective Newton constant evaluated at the horizon. With our boundary conditions we are in the grand-canonical ensemble, and the associated thermodynamic potential $W$ is related to the Euclidean action by $\mc S_E=\be W$. We can therefore extract the charges and thermodynamical quantities following the standard thermodynamical procedure,
\eq
S=\be^2\left.\frac{\p W}{\p\be}\right|_{\Phi,\Psi^{(i)} } =\frac{\mc A_h}{4G_h}\,,\quad
Q=-\left.\frac{\p W}{\p \Phi} \right|_{\be,\Psi^{(i)}  } =-\sigma\pi_A\,,\quad
Q_i=-\left.\frac{\p W}{\p\Psi_i}\right|_{\be,\Phi,\Psi^{(j)}_{j\neq i}}=-2\sigma\pi_{B_i}.
\label{entropy}\eeq
We see that the $\pi_A$ and $\pi_{B_i}$ are the charge densities carried by the Maxwell field and the three-form field. The mass of the black hole can be  obtained by Legendre transforming to the microcanonical ensemble,
\eq
M = W+TS+\Phi Q+\Psi_1Q_1+\Psi_2Q_2=-\frac{\sigma}{8\pi G}f_1.
\label{mass}\eeq
For variations along a family of solutions, we have $\delta\mc Q=0$, and \eqref{deltaQ} expresses the first law of thermodynamics \cite{Wald:1993ki},
\eq
\delta M = T\,\delta S +\Phi\,\delta Q+\Psi_1\,\delta Q_1+\Psi_2\,\delta Q_2.
\eeq
An important remark is that due to the non-minimal coupling between gravity and the scalar field $\phi$, the entropy of the black hole deviates from the Bekenstein-Hawking formula in that the gravitational coupling constant appearing in \eqref{entropy} is the effective Newton constant on the horizon $G_h$. The reason is that in the minimal frame the metric carries an extra conformal factor \eqref{bhminimal}, and the area of the horizon is then given by  $\mc A^{\text{min}}_h=(G/G_h)\si r_h^2$,  leading to the usual $\mc A_h^\text{min}/4G$ form for the entropy in that frame (see also the discussion in \cite{Martinez:2005di}).

We can now apply our results to the black holes of the previous section, starting with the hairy black hole \eqref{hairy}--\eqref{hairyfields}.
Its metric has $f_1=-2p^2G\mu$ and hence its mass \eqref{mass} is
\eq
M=\frac\si{4\pi}\,p^2\mu,
\eeq
Its temperature is obtained by requiring no conical singularity in the Euclidean metric at the horizon $r_h$ according to \eqref{conical},
\eq
T=\frac{V'(r_h)}{4\pi}=\frac{1}{\pi\ell^2}\left(r_h-\frac{|p|\ell}{2}\right),
\eeq
and its entropy is given by
\eq
S=\frac{\si r_h^2}{4G_{h}}
\eeq
The Maxwell and axionic charges can be read off the asymptotic expansion of the corresponding fields using \eqref{entropy},
\eq
Q=\frac{\si q}{4\pi},\qquad
Q_i=\frac{\si p}{\sqrt{8\pi G}}.
\eeq
The corresponding potentials are given by,
\eq
\Phi=\frac q{r_h},\qquad
\Psi_i=-\frac p{\sqrt{8\pi G}}\lp r_h+\frac{2\pi G}{3\al\ell^2}\frac{G^2\mu^2}{r_h+G\mu}\rp.
\eeq

The other solution competing in this ensemble is the undressed axionic Reissner-Nordstr\"om-AdS black hole \eqref{RN1}--\eqref{RN2}, for which the scalar field vanishes. This time $f_1=-2G\mu$ and its mass is
\eq
M=\frac\si{4\pi}\mu.
\eeq
Its temperature and entropy is
\eq
T=\frac1{2\pi\ell^2}\lp r_h+\frac{G\mu\ell^2}{r_h^2}-\frac{Gq^2\ell^2}{r_h^3}\rp,\qquad
S=\frac{\si r_h^2}{4G}.
\eeq
The entropy has the Bekenstein-Hawking form since $G_h=G$. Finally the charges and potentials are
\eq
Q=\frac{\si q}{4\pi},\qquad
Q_i=\frac{\si p}{\sqrt{8\pi G}},\qquad
\Phi=\frac q{r_h},\qquad
\Psi_i=-\frac{pr_h}{\sqrt{8\pi G}}.
\eeq

We conclude this section with one last note on the solutions with constant $\phi$ given in \eqref{phiRN1}--\eqref{phiRN2}. It would be tempting to weaken the boundary conditions on the fields to include them into the analysis. However, doing so, one would have to impose Dirichlet boundary conditions on the scalar field thus changing the thermodynamical ensemble.

\section{Phase diagram and superconducting phase transition\label{sec::phase}}

With the results of the previous section we can tackle the study of the phases of these black holes. For simplicity, we will restrict to an ensemble where $(T,\Phi,Q_i)$ are kept constant. The relevant thermodynamical potential is Gibbs' free energy, obtained by Legendre transforming $W$,
\eq
\mc G(T,\Phi,Q_1,Q_2)=W+\Psi_1 Q_1+\Psi_2 Q_2=M-TS-\Phi Q.
\eeq
Following \cite{Martinez:2010ti}, it is convenient to introduce the inverse coupling constant $a$,
\eq
a=\frac{2\pi G}{3\al\ell^2},\qquad
a\geq1,
\eeq
and to work with the dimensionless quantities defined by\footnote{The charges $Q_1$ and $Q_2$ are not independent in this ensemble; for the solutions we have, they must be equal, $Q_1=Q_2$, and therefore we have simply $\varpi=|p|$. We will generically denote them as $Q_i$.},
\eq
\varpi=\frac{\sqrt{4\pi G}}{\si}\sqrt{Q_1^2+Q_2^2}\,,\qquad
\xi=\frac{2\pi\ell T}{\varpi}.
\eeq
In these variables, regularity condition \eqref{ghost} for the hairy solution \eqref{hairy} imposes on $\xi$ the bounds
\eq
\frac{\sqrt a-1}{\sqrt a+1}<\xi<\frac{\sqrt a+1}{\sqrt a-1}.
\label{xireg}\eeq
Moreover, the hairy black hole only exists for parameters that satisfy \eqref{q}. This relation, written in terms of the thermodynamical variables of the ensemble we are interested in, reads
\eq
\Phi^2=\frac{\varpi^2}{4G}(a-1)(\xi-1)^2,
\label{consphi}\eeq
and can be used to eliminate the Maxwell potential.
Some simple algebra shows then that the Gibbs free energy of the hairy black hole is given by,
\eq
\mc G_{\rm hairy}=-\frac{\si\ell\varpi^3}{32\pi G}
\lp (\xi+1)^2+a(\xi-1)^2
\rp.
\eeq
The analogous computation for the axionic Reissner-Nordstr\"om-AdS black hole, restricting to the solutions satisfying the constraint \eqref{consphi}, yields,
\eq
\mc G_{\rm RN}=-\frac{\si\ell\varpi^3}{108\pi G}\lb
\xi^3+\frac{9\xi}{2}\lp1+\frac14(a-1)(\xi-1)^2\rp+\lp\xi^2+3\lp1+\frac14(a-1)(\xi-1)^2\rp\rp^{3/2}\rb.
\eeq
\begin{figure}[tb]
\centerline{\includegraphics{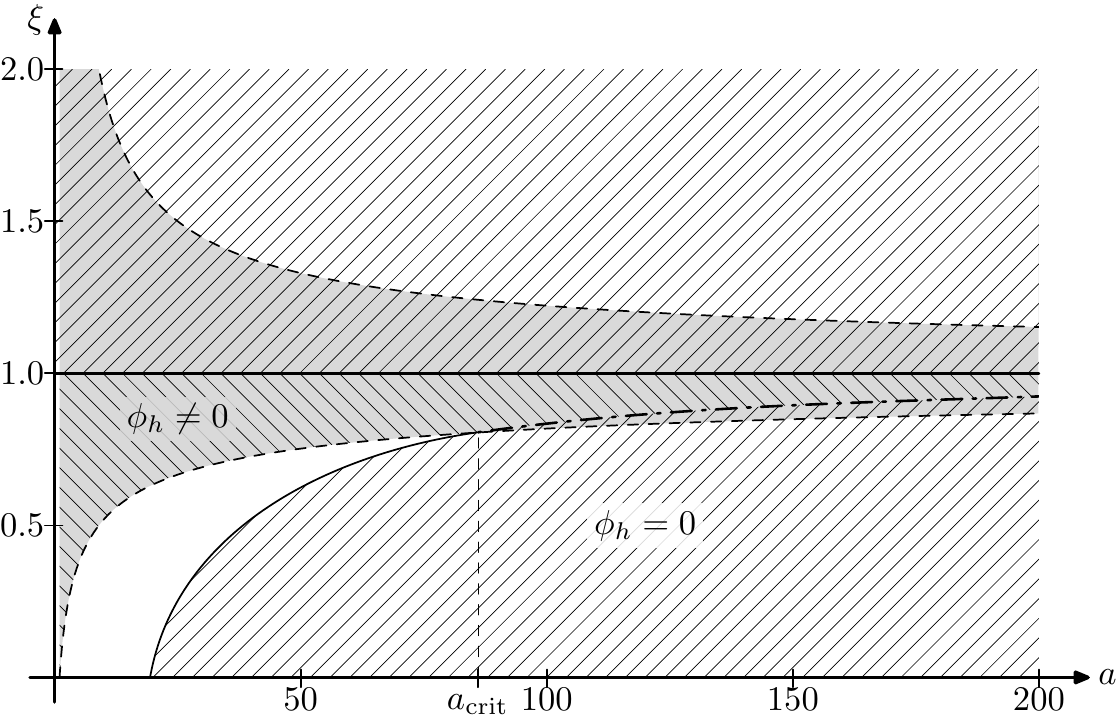}}
\caption[]{{\sc Coupling constant dependence of the phases.} In the shaded region, regular hairy black holes exist and compete with the axionic RN-AdS black hole. They dominate in the hatched region
\protect\raisebox{-2pt}{\includegraphics[width=1.8em]{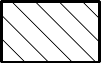}}, while the axionic RN-AdS black holes dominate the other hatched region \protect\raisebox{-2pt}{\includegraphics[width=1.8em]{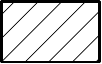}}. They are separated by a second order phase transition (continuous line at $\xi=1$) and at lower temperatures by a first order phase transition (dashed dotted line). In the white region, singular hairy solutions have lower free energy and should dominate.}
\label{fig1}\end{figure}
\begin{figure}[tb]
\centerline{\quad\includegraphics{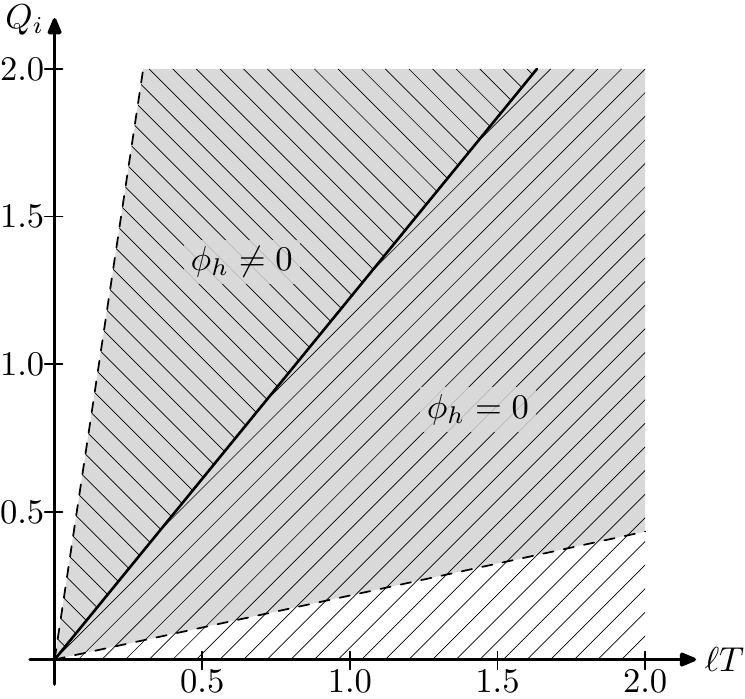}
\hfill\includegraphics{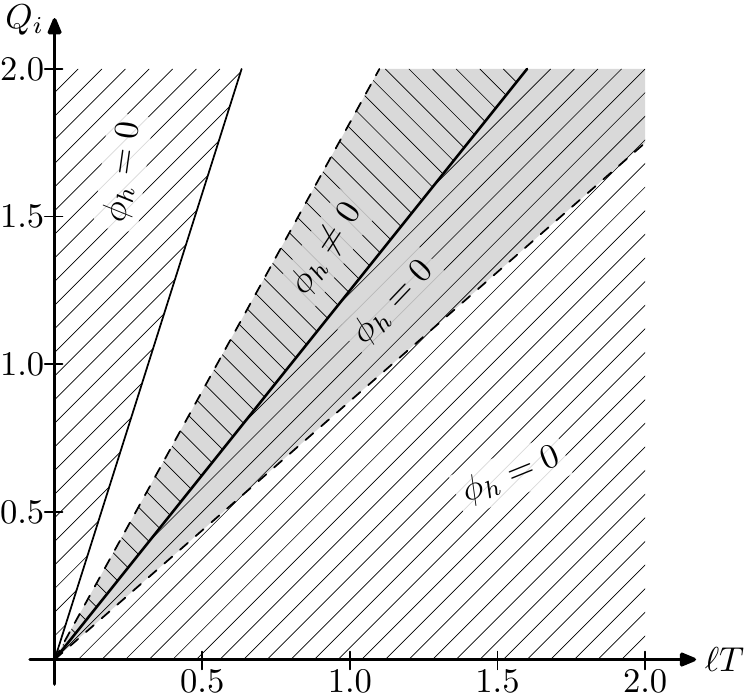}\quad}
\caption[]{{\sc Phase diagrams in $(T,Q_i)$ plane.} For low $a$, as the temperature is lowered one goes from an axionic RN-AdS dominated phase to a hairy phase, through a second order phase transition (thick continuous line in the diagrams), and the order parameter $\phi_h$ acquires a non-vanishing value in the process. Lowering the temperature further, the hairy black holes become singular while still having lower free energy than the axionic RN-AdS black hole (white region). This is exemplified for $a=2$ in the left diagram, and it is valid for $1<a<3 (3 + 2\sqrt{3})$. For $3 (3 + 2\sqrt{3})<a<a_{\rm crit}$, with $a_{\rm crit}\approx86$, at very low temperatures one finds an additional bald phase with $\phi_h=0$. As a representative of this situation, the $a=30$ case is depicted on the right figure. In the diagrams, the shaded region indicates the region where regular hairy black holes and axionic RN-AdS black holes coexist, the hatched region
\protect\raisebox{-2pt}{\includegraphics[width=1.8em]{hairy}} is the hairy $\phi_h\neq0$ phase, and the other hatched region \protect\raisebox{-2pt}{\includegraphics[width=1.8em]{rn}} corresponds to the axionic RN-AdS phase with $\phi_h=0$.}
\label{fig2}\end{figure}
\begin{figure}[htb]
\centerline{\includegraphics{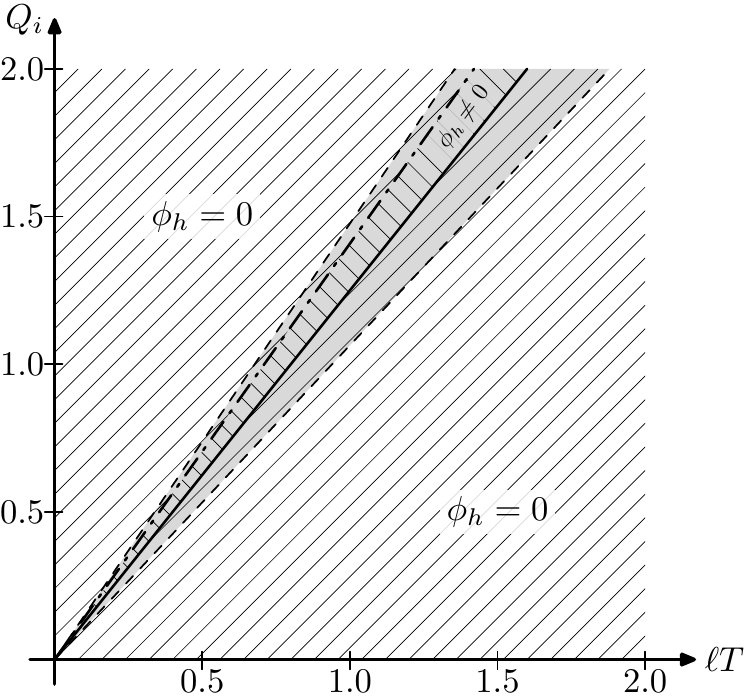}}
\caption[]{{\sc Phase diagram in $(T,Q_i)$ plane for small quartic coupling $\al$ ($a>a_{\rm crit}$).} In this case, every point of the phase diagram is dominated by a regular black hole: the axionic RN-AdS black hole in the hatched region \protect\raisebox{-2pt}{\includegraphics[width=1.8em]{rn}}, and the hairy black hole in the hatched region \protect\raisebox{-2pt}{\includegraphics[width=1.8em]{hairy}}. The higher temperature phase transition is second order (continuous line) and the lower temperature phase transition is first order (dashed dotted line). In the shaded region both phases coexist. The diagram portrays the particular case with $a=150$.}
\label{fig3}\end{figure}

Note that the axionic charge $\varpi$ appears explicitly as a $\varpi^3$ overall factor only, in both $\mc G_{\rm hairy}$ and $\mc G_{\rm RN}$. Hence, in this ensemble, the black hole phase depends on the single dimensionless variable $\xi$, and is determined by the sign of $\Delta\mc G=\mc G_{\rm RN} - \mc G_{\rm hairy}$. The solution with lowest free energy will be thermodynamically favoured. An interesting observation is that setting $\varpi=1$, the free energy difference $\Delta\mc G$ reduces to the free energy difference between the {\em hyperbolic} Reissner-Nordstr\"om-AdS and the {\em hyperbolic} hairy black hole, as computed in \cite{Martinez:2010ti}. The reason is that the axionic fields effectively change the curvature of the horizon (see \cite{Bardoux:2012aw}), and the functions $f(r)$ coincide for the planar axionic solutions with $\varpi=1$ and the hyperbolic non-axionic solutions.
Hence, the thermodynamic properties, and in particular the phases, of the $\varpi=1$ black holes, are the same as the properties of hyperbolic black holes obtained in \cite{Martinez:2010ti}. Therefore, since the explicit dependence on $\varpi$ is only through a common factor of $\varpi^3$ in the free energies, for any other value $\varpi\neq1$, the thermodynamic properties can be deduced by rescaling $\xi$ (or equivalently, the temperature) accordingly.

First, notice that $\Delta\mc G=0$ for $\xi=1$. This corresponds to the critical temperature
\eq
T_{\rm crit}=\sqrt{\frac G\pi}\frac{\sqrt{Q_1^2+Q_2^2}}{\si\ell},
\eeq
where a second order phase transition takes place. For $T>T_{\rm crit}$ the axionic RN-AdS black hole dominates and the scalar field vanishes. Lowering the temperature, as one crosses the critical temperature, the scalar field is spontaneously generated, acquiring a finite value $\phi_h$ on the horizon of the black hole: for $T<T_{\rm crit}$ the hairy black hole dominates, and the value $\phi_h$ of the scalar field on the horizon can be taken as the order parameter of the transition (see Figure~\ref{fig1}).

As one lowers the temperature further, a second  $\Delta\mc G=0$ curve appears when $a>3(3+2\sqrt{3})$, that signals that the thermodynamically favoured state is again a scalar-free RN-AdS black hole. As long as $a>a_{\rm crit}$, with $a_{\rm crit}\approx86$ \cite{Martinez:2010ti}, this is a genuine first order phase transition. However, for lower values of the quartic scalar coupling, the hairy phase is not regular, since it is outside the range \eqref{xireg}, and it is not clear what the correct equilibrium state is in that region; it might simply not make sense to try to extrapolate the physics to strongly coupled regimes for which the irregular hairy solution has a diverging effective Newton constant outside the horizon.

The phase diagrams are summarised in Figures~\ref{fig2} and \ref{fig3} for the different ranges of the coupling constant in the $(T,Q_i)$ plane, $Q_i$ being the axionic charge.

\section{Conclusions and some thoughts on holographic superconductors \label{conclusion}}
In this paper we have presented hairy black hole solutions with locally AdS asymptotics, charged with axionic and electromagnetic fields. The solutions have a conformally or minimally coupled scalar field and belong to the class of MTZ black holes \cite{Martinez:2002ru,Martinez:2004nb}. The most  important characteristic of the solutions found in this article, is that they are the planar versions of the MTZ black holes. The only known planar solution with a  conformally coupled scalar field in four dimensions was a pure AdS stealth solution \cite{Martinez:2002ru}. Here, two axions are necessary in order to keep the planar two-dimensional horizon homogeneous. Starting with the BBMB black hole \cite{Bekenstein:1974sf,Bekenstein:1975ts,Bocharova:1970}, Mart\'\i nez, Troncoso and Zanelli demonstrated \cite{Martinez:2002ru,Martinez:2004nb} that one has to consider a cosmological constant with a self-interaction potential in order for the scalar field to be well behaved at the constant---and non zero curvature---event horizon. Here, in addition, the axionic scalars are needed in order to provide the curvature scale necessary for a planar---rather than curved---black hole horizon. In this sense the double axionic charge is necessary and not an ad hoc charge for the solutions. Similarly if one were to look for the $D$ dimensional version of these black holes then in principle we would need $D-2$ axion fields or equivalently $D-2$ in number, $(D-1)$-form field strengths \cite{Bardoux:2012aw}. The basic as well as the thermodynamic properties of the solutions are similar to the hyperbolic version of the MTZ black holes \cite{Martinez:2005di,Martinez:2010ti}. They can have a negative mass with a triple horizon geometry and they compete with the axionic RN-AdS black hole in the grand-canonical ensemble. As the temperature is lowered and reaches the critical temperature, the axionic RN-AdS black hole undergoes a second order phase transition to the hairy black hole, and get spontaneously dressed by the scalar field.

Black holes are thermodynamic objects and respond to perturbations with calculable dynamical transport coefficients.
Hairy planar black holes present a particular interest since they are possible gravitational duals to holographic superconductors in two spatial dimensions \cite{Koutsoumbas:2009pa,Myung:2010rb}. The idea (or hope) is that holographic superconductors may describe phases of certain layered materials which are high temperature superconductors such as certain copper oxides (cuprates). Cuprates are superconductors at up to $134\,{\rm K}$ at atmospheric pressure, which obviously makes them particularly interesting for technological applications. High temperature involves strong coupling and thus their superconductivity (pairing properties, dynamical transport coefficients, etc.) are poorly understood theoretically \cite{Horowitz:2010gk,Hartnoll:2009sz,Hartnoll:2008vx,Hartnoll:2008kx,Herzog:2009xv}. In their most basic bottom-up description, holographic superconductors are described by a charged complex scalar field coupled to an electromagnetic field and to Einstein-Hilbert gravity with a cosmological constant. The Langrangian reads,
\beq
\mc L=R+\frac{6}{\ell^2}-\frac{1}{4}F^2-|\nabla \Phi-i Q A \Phi|^2-V(|\Phi|).
\eeq
Here $Q$ plays the role of charge or primary hair associated to the scalar field. In a strongly coupled theory one expects to have multiple condensates of different operators and not just the one as pictured here. Clearly the above action is a simple toy model (see e.g.~\cite{Hartnoll:2009sz}, Section 4.2) where the charged scalar plays the role of the condensate and, in the dual picture, the black hole hair.
In a similar way to the Higgs mechanism in particle physics \cite{Gubser:2008px}, as we lower the temperature, and for large enough charge $Q$, the scalar complex field destabilizes from its trivial constant value and a hairy black hole becomes thermodynamically preferred to the bald RN-AdS planar black hole. The hairy black hole resulting from \eqref{action} has similarities and certain differences to the prototype picture given above. Clearly a difference is the presence of the axions and the fact that we have non minimally coupled scalars. The presence of a real rather than complex scalar playing the role of the condensate, is however consistent with the above action. In fact it is easy to see that for a given electric or magnetic field in a static metric ansatz the phase of the complex scalar is zero and thus the scalar can be taken consistently as a real field \cite{Horowitz:2010gk}.  More importantly our scalar here is not charged, $Q=0$, and therefore it is not a Higgs mechanism that provides the transition between the two black holes. It would seem that this is due to the non minimal coupling to gravity and the presence of the axionic charge (see also \cite{Koutsoumbas:2009pa}). It is maybe here that there is an indication that the axions present are part of a phase for a complex uncharged scalar field just as is the case of global string topological defects \cite{cosmic,Davis:1988rw,Vilenkin:1986ku}. A quick check shows however that the precise coupling function at hand \eqref{action} is not the right one, and we would still have to explain the presence of the second axion. The stringy origin of the fields into play is an interesting route to pursue since the presence of the axions is necessary for the hairy black hole to exist and should therefore be a necessary ingredient for this type of holographic superconductor if this construction is to make any sense.

Furthermore, this work opens up novel possibilities for finding other hairy black holes, beyond the standard conformal coupling and also in higher dimensions \cite{bcch}. In fact, planar black holes have one less curvature scale associated to them, and often, as a result, enjoy better integrability properties than non-planar black holes. Hence, given that we now have the key to their construction, a novel start in this direction is possible. Coupling functions and potentials, as appearing in \eqref{action}, will then change providing a more general map for  holographic applications. Finally, new classes of hairy black hole solutions were recently found \cite{Kolyvaris:2009pc,Kolyvaris:2011fk,Anabalon:2012tu,Anabalon:2012ta}
and it would also be fruitful to pursue work in this direction, for example to find planar C-metric versions of the MTZ black holes \cite{Charmousis:2009cm,Anabalon:2009qt}.

\section*{Acknowledgements}
It is a pleasure to thank B.~Gout\'eraux, R.~Gregory,  F.~Nitti and R.~Parentani for useful discussions. We thank in particular M.~Hassa\"\i ne, for numerous discussions, comments and for reading through the manuscript.
This work was partially supported by the ANR grant STR-COSMO, ANR-09-BLAN-0157.
MMC was additionally partially supported by
the ERC Advanced Grant 226371,
the ITN programme PITN-GA-2009-237920,
the IFCPAR CEFIPRA programme 4104-2 and the ANR programme NT09-573739 ``string
cosmo''.

\end{document}